\documentclass{article}
\usepackage[utf8]{inputenc}
\usepackage{microtype}
\DisableLigatures{encoding = *, family = * }
\usepackage{amsmath}
\usepackage{algorithm}
\usepackage{algpseudocode}
\usepackage{algpascal}

\usepackage{pdflscape}

\usepackage{algorithm}
\usepackage{algpseudocode}
\usepackage{algpascal}
\usepackage{subfig}
\usepackage{float}
\usepackage{makeidx}
\usepackage{balance}
\usepackage{graphicx}
\usepackage{lscape}
\usepackage{multirow}
\usepackage{multicol}
\usepackage{breqn}
\usepackage{verbatim}
\usepackage{newclude}
\usepackage{array}
\usepackage{booktabs}
\usepackage{graphicx}
\usepackage{caption}
\usepackage{acronym}

\usepackage{tablefootnote}
\usepackage{threeparttable}

\usepackage[table]{xcolor}
\definecolor{lightgray}{gray}{0.96}
\usepackage{pgfplots}
\pgfplotsset{compat=newest}
\usetikzlibrary{plotmarks}

\newlength\figureheight
\newlength\figurewidth
\acrodef{HMM}{hidden Markov model}
\acrodef{FHMM}{fractional hidden Markov model}
\acrodef{NILM}{Non-intrusive load monitoring}
\acrodef{PF}{particle filtering}
\acrodef{ETC}{effort-to-compress}
\acrodef{NSRPS}{non-sequential recursive pair substitution}
\acrodef{PC}{power similarity complexity}
\acrodef{C}{combined complexity}
\acrodef{ACC}{accuracy}
\acrodef{NIOM}{Non-Intrusive Occupancy Monitoring}
\acrodef{HEMS}{Home Energy Management System}
\acrodef{HVAC}{Heating ventilation and air conditioning}
\acrodef{BN}{Bayesian network}
\acrodef{EM}{expectation maximization}
\acrodef{JT}{junction tree}
\acrodef{ACC}{accuracy}

\usepackage{siunitx}
\usepackage{tikz}
\usepackage{pgfplots}
\pgfplotsset{width=7cm,compat=newest}

\begin{document}
\def\figscale{.92}
\setlength{\baselineskip}{.97\baselineskip}

\title{GREEND: An Energy Consumption Dataset of Households in Italy and Austria}

\author{Andrea Monacchi,
\thanks{We want to thank Benjamin Steinwender, Micha Rappaport and Manfred P\"{o}echacker for the support provided while carrying out the measurement campaign.
The work of Monacchi, Egarter and Elmenreich is supported by Lakeside Labs, Klagenfurt, Austria and funded by the European Regional Development Fund (ERDF) and the Carinthian Economic Promotion Fund (KWF) under grant KWF 20214 — 23743 — 35469 u. 35470. The work of D'Alessandro and Tonello is cofunded by the Interreg IV Italy-Austria ID-6462 MONERGY project.} Dominik Egarter, Wilfried Elmenreich\\
\small Institute of Networked and Embedded Systems / Lakeside Labs,\\[-0.8ex] 
\small Alpen-Adria-Universit\"at Klagenfurt, Austria\\[-0.8ex]
\small \texttt{name.surname@aau.at}\\
\and
Salvatore D'Alessandro, Andrea M. Tonello\\
\small WiTiKee s.r.l.\\[-0.8ex]
\small via Duchi d'Aosta 2, Udine, Italy\\[-0.8ex]
\small \texttt{surname@witikee.com}
}
\date{}

\maketitle

\begin{abstract}
Home energy management systems can be used to monitor and optimize consumption and local production from renewable energy.
To assess solutions before their deployment, researchers and designers of those systems demand for energy consumption datasets.
In this paper, we present the GREEND dataset, containing detailed power usage information obtained through a measurement campaign in households in Austria and Italy.
We provide a description of consumption scenarios and discuss design choices for the sensing infrastructure.
Finally, we benchmark the dataset with state-of-the-art techniques in load disaggregation, occupancy detection and appliance usage mining.\\[0.8em]
\end{abstract}

\section{Introduction} \label{sec:introduction}
Stability issues arise from the progressive installation of renewable energy generation and the diffusion of electric vehicles.
Demand response exploits a price signal to reflect fluctuations in the availability of energy in the grid.
This incentives the coordination of electrical devices in order to optimize running costs,
fostering awareness to increase conservation and efficiency \cite{CarrieArmel2013,monacchi:2013Nov}
or to automatically schedule the operation of selected devices to off-peak periods \cite{palensky}.
Nevertheless, control strategies might somehow disrupt user's daily life routines.
Collecting information of reoccurring activities can increase the effectiveness of control strategies,
as they can consider necessities of inhabitants to minimize the discomfort produced.
Problems such as load detection and occupancy modeling for HVAC 
optimization can be solved using well-established data mining and machine learning techniques, which demand suitable datasets.

In order to enable research on energy and sustainability problems, it is necessary to build upon publicly available data in order to elaborate solutions that work in the real world.  
In this paper we discuss available energy consumption datasets and present a new publicly available dataset that we name GREEND\footnote{GREEND: GREEND Electrical ENergy Dataset.}, containing power usage information at device level being obtained through a measurement campaign in households in Austria and Italy. 
We describe the hardware and software setup of the measurement campaign and elaborate the modeling of loads for energy management applications. 
The paper focuses on the dataset and it is meant to provide metadata describing the involved consumption scenarios.
Along with the dataset we are also releasing our code base for the measurement campaign as a Sourceforge project.
Moreover, we benchmark the dataset by presenting experiences with load disaggregation, occupancy detection and appliance usage mining using state-of-the-art techniques.

The remainder of the paper is as follows.
To identify the main requirements of an energy consumption dataset, we survey existing datasets in Section \ref{sec:datasets}.
In Section \ref{sec:datgreend}, we report the design of the measurement campaign.
We describe our measurement infrastructure and provide a characterization of the deployed scenarios, in terms of both residents and monitored electrical devices.
In Section \ref{sec:nilm}, we apply a state-of-the art load disaggregation technique based on particle filtering to show a possible application of the dataset.
Similarly, in section \ref{sec:occupancy} and \ref{sec:usage} we apply the dataset respectively to observe occupancy patterns and mine appliance usage patterns.
Section \ref{sec:conclusion} concludes the paper and anticipates future directions.
\section{Other consumption datasets} \label{sec:datasets}
To overcome the previously introduced challenges, researchers and designers of \ac{HEMS} require consumption datasets where solutions can be assessed beforehand.
A widely used dataset in the load disaggregation community is the reference dataset (REDD), publically released by MIT in 2011 \cite{kolter-kdd-2011}.
Moved by a similar aim, various other datasets have been shared.
As shown in Table \ref{tab:datasets}, the main classification attributes are sampling frequency and characteristics of the signal being measured, active power (P), reactive power (Q), apparent power (S), energy (E), frequency (f), phase angle ($\Phi$), voltage (V) and current (I).
Certain datasets, such as REDD and BLUED \cite{anderson_blued:_2012} monitor a small number of households at a high sampling frequency.
This depends on the requirements of load disaggregation, where a higher frequency allows for extraction of more representative features capturing the transient behavior \cite{CarrieArmel2013}.
As noticeable in the table, most of datasets were collected in the USA, under a $120V$ voltage while European countries commonly work under $230V$.
Moreover, certain peculiarities such as the weather and climate depend on the location of the measurement campaign.
To the best of our knowledge, GREEND is the first 1Hz consumption dataset for Austria and Italy.
The type and number of devices, as well as the number of monitored households significantly constraints the final application of the dataset.
A statistical analysis of consumption behavior of residents would require a high number of households, such as in HES and OCTES.
Besides, seasonal consumption behaviors cannot be captured by a short-term measurement campaign, for instance lasting days or months.
Also, some datasets monitor different households over different time windows, which makes a comparison of the dwellings impossible. 
Moreover, the context under which appliances are used over the day is an essential factor determining the complexity of the demand.
Therefore, collection should take place in real environments and not in a lab or on a device testbench.
For example TraceBase \cite{reinhardt12tracebase} and ACS-F1 \cite{ridi} provide consumption tracks from appliances collected in households and offices.
While they constitutes a great source of device signatures, they currently do not provide information of consumption scenarios, which results in the impossibility to analyze device interdependence.
GREEND is meant to overcome these limitations as it will be explained in the following section.

\begin{landscape}
\begin{table*}[bt]
 \centering
 \caption{Existing datasets for energy consumption in households}\label{tab:datasets}
\scalebox{0.8}{
\rowcolors{1}{}{lightgray}
	\centering
	\begin{threeparttable}[b]
    \begin{tabular}[c]{| l | p{10em} | p{10em} | p{10em} | p{10em} | p{10em} | p{10em} |}
    \hline
    Dataset																				&		Location					&	Duration									& \#Houses					&	\#Sensors (per house)						&    Features									& Resolution							\\
    \hline
    ACS-F1 \cite{ridi}																	&		Switzerland					&	1 hour session (2 sessions)					&	N/A						&	100 devices in total (10 types)				&	I, V, Q, f, $\Phi$							&	10 secs								\\			
    AMPds \cite{makonin2013ampds}														&		Greater Vancouver			&	1 year										&	1						&	19											&	I, V, pf, F, P, Q, S						&	1 min							\\
    BLUED \cite{anderson_blued:_2012}													&		Pittsburg, PA				&	8 days										&	1						&	Aggregated									&	I, V, switch events							&	12 Khz							\\
	\textbf{GREEND}																		&		Austria, Italy				&	1 year	(3-6 months completed)				& 	9						&	9 											& 	P											&	1 Hz							\\
	HES																					&		UK							&	1 month (255 houses) - 1 year (26 houses)	&	251						&	13-51										&	P											&	2 min							\\
    iAWE \cite{Batra:2013}																&		India						&	73 days										&	1						&	33 sensors (10 appliance level)				&	V, I, f, P, S, E, $\Phi$						&	1 Hz							\\
    IHEPCDS\tnote{1}
    																					&		France						&	4 years										&	1						&	3 circuits									&	I, V, P, Q									&	1 min							\\
    OCTES\tnote{2}
    																					&		Finland, Iceland, Scotland	&	4-13 months									&	33						&	Aggregated									&	P, Energy price								&	7 secs							\\
	REDD \cite{kolter-kdd-2011}															&		Boston, MA					&	3 - 19 days									&	6						&	9-24										&	Aggregate: V, P; Sub-metered: P				&	15 Khz (aggr.), 3 sec (sub)		\\
	Sample dataset\tnote{3}
																						&		Austin, TX					&	7 days										&	10						&	12											&	S											&	1 min							\\
    Smart* \cite{barker2012smart}														&		Western Massachussets		&	3 months									&	1 Sub-metered +2 (Aggregated + Sub-metered)		&	25 circuits, 29 appliance monitors			&	P, S (circuits), P (sub-metered)			&	1 Hz							\\
    Tracebase \cite{reinhardt12tracebase}												&		Germany						&	N/A											&	15						&	158 devices in total (43 types)				&	P											&	1-10 sec						\\
    UK-DALE \cite{2014arXiv1404.0284K}													&		UK							&	499 days									&	4						&	5 (house 3) - 53 (house 1)					&	Aggregated P, Sub P, switch-status			&	16 Khz (aggr.), 6 sec (sub.)	\\
    \hline
    \end{tabular}
    \begin{tablenotes}
    	\item[1] http://tinyurl.com/IHEPCDS
    	\item[2] http://octes.oamk.fi/final/
    	\item[3] http://www.pecanstreet.org/projects/consortium/
  	\end{tablenotes}
  	\end{threeparttable}
}
\end{table*}
\end{landscape}
\section{Dataset for Italy and Austria}\label{sec:datgreend}
\subsection{The consumption dataset}
The measurement campaign is carried out within the MONERGY\footnote{http://www.monergy-project.eu} project, in which we aim at proposing solutions to reduce energy consumption in the Austrian (AT) region of Carinthia and the
Italian (IT) region of Friuli-Venezia Giulia. In particular, we identified the following requirements:
\begin{itemize}
  \item \textit{Features:}
  For the selection of the features of interest we considered the requirements of load disaggregation applications,
  as they are generally stricter than user and appliance modeling \cite{Zeifman2012}.
  Accordingly, we decided to collect active power measurements at 1Hz, as this allows the identification of more than 8 devices through load disaggregation algorithms \cite{CarrieArmel2013}.
  Each entry is associated to a UTC Unix timestamp so that measurements can be matched to contextual data such as weather.
  \item \textit{Device selection:}
  The selection of devices followed the energy hogs identified in \cite{monacchi:2013Nov}.
  We favored diversity in the dataset to promote multiple applications.
  However, we faced issues in accessing certain devices in Austria.
  For instance, electric boilers are commonly connected to a separated meter to be charged with a reduced energy tariff.
  \item \textit{Household selection:}
  The selection of householders was driven by the findings identified in \cite{monacchi:2013Nov}.
  In particular, we wanted to promote diversity of scenarios, for instance involving different types of dwellings and consumers.
  A more detailed description of the scenarios is reported in Section \ref{subsec:deployment}.
  \item \textit{Campaign duration:}
  The campaign was designed to last one year, in order to observe and be able to model seasonal consumption behavior of inhabitants.
  The first household is being monitored (\#0) since the end of December 2013.
  Most of other households followed in January 2014, while the last two platforms (house \#7 and \#8) were deployed in April.
\end{itemize}
\subsection{Deployment}\label{subsec:deployment}
At the time of writing, we are collecting consumption data in the following scenarios:
\begin{itemize}
  \item \textit{House \#0} a detached house with 2 floors in Spittal an der Drau (AT).
  The residents are a retired couple, spending most of time at home.
  \item \textit{House \#1} a detached house with 2 floors in Villach (AT).
  The residents are 3 university students, having irregular working-like days.
  \item \textit{House \#2} an apartment with 1 floor in Klagenfurt (AT).
  The residents are a young couple, spending most of daylight time at work during weekdays, mostly being at home in evenings and weekend.
  \item \textit{House \#3} a detached house with 2 floors in Spittal an der Drau (AT).
  The residents are a mature couple (1 housewife and 1 employed) and an employed adult son (28 years).
  \item \textit{House \#4} a detached house with 2 floors in Klagenfurt (AT).
  The residents are a mature couple (1 working part-time and 1 full time), living with two young kids.
  \item \textit{House \#5} an apartment with 2 floors in Udine (IT).
  The residents are a young couple, spending most of daylight time at work during weekdays, although being at home in evenings and weekend.
  \item \textit{House \#6} a detached house with 2 floors in Colloredo di Prato (IT).
  The residents are a mature couple (1 housewife and 1 employed) and an employed adult son (30 years).
  \item \textit{House \#7} a terraced house with 3 floors in Udine, (IT).
  The residents are a mature couple (1 working part-time and 1 full time), living with two young children.
  \item \textit{House \#8} a detached house with 2 floors in Basiliano (IT).
  The residents are a retired couple, spending most of time at home.
\end{itemize}
The device configurations for the selected households are shown in Table \ref{tab:devs}.
\begin{table}
 \centering
 \caption{Device configurations in the monitored households}\label{tab:devs}
 \rowcolors{1}{}{lightgray}
    \begin{tabular}{| c  m{0.8\columnwidth} |}
    \hline
    \textbf{House}		&	\textbf{Devices}\\
    \hline
    	0		&	Coffee machine, washing machine, radio, water kettle, fridge w/ freezer, dishwasher, kitchen lamp, TV, vacuum cleaner\\
		1		&	Radio, freezer, dishwasher, fridge, washing machine, water kettle, blender, network router\\
		2		&	Fridge, dishwasher, microwave, water kettle, washing machine, radio w/ amplifier, dryier, kitchenware (mixer and fruit juicer), bedside light\\
		3		&	TV, NAS, washing machine, drier, dishwasher, notebook, kitchenware, coffee	machine, bread machine\\
		4		&	Entrance outlet, Dishwasher, water kettle, fridge w/o freezer, washing machine, hairdrier, computer, coffee machine, TV\\
		5		&	Total outlets, total lights, kitchen TV, living room TV, fridge w/ freezer, electric oven, computer w/ scanner and printer, washing machine, hood\\
		6		&	Plasma TV, lamp, toaster, hob, iron, computer w/ scanner and printer, LCD TV, washing machine, fridge w/ freezer\\
		7		&	Hair dryer, washing machine, videogame console and radio, dryer, TV w/ decoder and computer in living room, kitchen TV, dishwasher, total outlets, total lights\\
		8		&	Kitchen TV, dishwasher, living room TV, desktop computer w/ screen, washing machine, bedroom TV, total outlets, total lights\\
    \hline
    \end{tabular}
\end{table}

\subsection{Measurement setting}
\begin{figure}[h!]
	\centering
	\includegraphics[trim=7.7cm 23.1cm 3.9cm 3.2cm, clip, width=0.85\columnwidth]{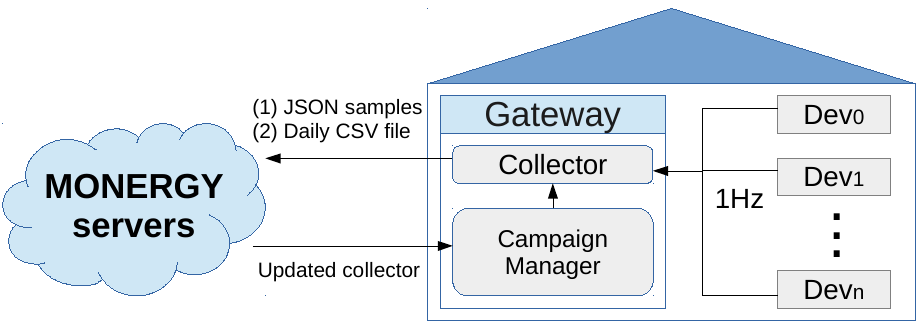}
	\caption{The sensing infrastructure}
	\label{fig:infrastructure}
\end{figure}

\begin{figure}[h!]
	\centering
	\includegraphics[width=0.8\columnwidth]{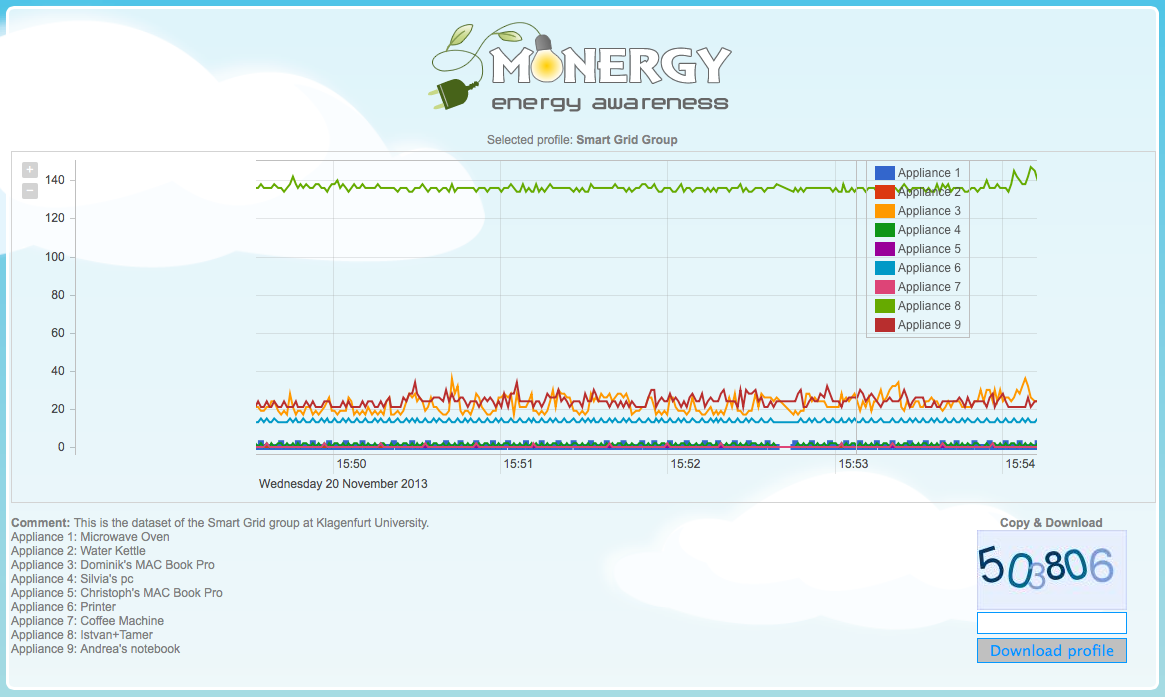}
	\caption{The website of the testbed}
	\label{fig:testbed}
\end{figure}

Our deployment consists of an ARM-based platform (e.g., Raspberry Pi or BeagleBone) connected to an Anker Astro E5 15000mAh external battery\footnote{http://www.ianker.com/support-c1-g228.html},
as well as a Plugwise Basic kit\footnote{http://www.plugwise.com}.
The Plugwise kit consists of a Zigbee network of 9 sensing outlets, each collecting active power measurements from the connected load.
The collection takes place in epochs. At each epoch, the system collects power measurements from each node and sleeps for the remaining time.
To provide periodicity, in presence of failures such as mispelled packets, nodes are skipped and retried at the end of the epoch if there is time.
In addition, a backoff time is used to manage faults, such as the ones resulting from temporary erroneous states or disconnection of nodes from the network.
The developed daemon uses the open source python-plugwise library and it is freely available both as a sourcecode and as a ready-to-use SD image\footnote{http://sourceforge.net/projects/monergy}.
The daemon is based on a collector script and a manager script. The manager is started at boot up by crontab.
Its task is to make sure that a collection script is running and that the running version is the latest available.
To this end, the manager periodically checks (default is 5 hours) the presence of new versions on the Monergy servers,
and if needed, it replaces the current version with the newest available.
In this way we can push updates to all households without needing to access each and every unit (see Fig.\ref{fig:infrastructure}).
As for the data storage, the current version of the daemon implements 4 different modalities:
i) local storage as a daily comma separated value (CSV) file,
ii) remote storage on a mysql server with visualization and quick download (See
Fig. \ref{fig:testbed}), iii) a combination of i and ii for double backup,
iv) daily csv file uploaded via sftp to the Monergy server (selected method).
\section{Case studies} \label{sec:experiments}
The introduction of smart metering produces fine-grained energy consumption data
allowing for the extraction of more valuable information of energy production and demand compared to having only an energy balance over an extended period of time.
Energy management systems (EMS) can be used to better manage local consumption and production from renewable sources.
Beside providing consumers with feedback regarding used appliances and their operational costs, this might also support appliance scheduling in order to minimize costs.
In this section, we report 3 case studies dealing with consumption information: i) load disaggregation, ii) occupancy detection, and iii) appliance usage modeling.
\subsection{Non-Intrusive appliance load monitoring} \label{sec:nilm}
\ac{NILM}, also known as load disaggregation, is the problem to detect and to classify appliances from the
total household power draw, based on specific characteristics of electrical devices.
Specifically, \ac{NILM} tries to solve the disaggregation of aggregated appliance power loads under the influence of noise in measurements and models.
\ac{NILM} was initially presented by Hart \cite{Hart1992}, although various approaches were later presented.
It is generally possible to distinguish between supervised and unsupervised techniques, depending on the necessity of labeled data to train the classifier.
Nevertheless, there exists no \ac{NILM} algorithm able to solve the problem in all its aspects.

Our evaluation is based on the load disaggregator presented in \cite{Egarter2013BuildSys}, which uses \ac{PF} to estimate the appliance state of all used appliances.
In detail, appliances are model as a \ac{HMM}, whose states define device states and are associated to the respective power demand.
Furthermore, the appliance models are combined into a \ac{FHMM} modeling the total household power demand.
The \ac{PF} is used to estimate the household power demand according to the given appliance models.
Thereafter, a simple decision marker determines the appliance condition and operational state,
based on thresholding and the given appliance models.
For the evaluation we used the data of house with ID \#0 and \#2 for $7$ consecutive days.
As the \ac{PF} is mainly dependent the number of used particles, we empirically identified $1000$ as the appropriate number.
The aggregated power draw is composed by 6 different appliances which are listed in Table \ref{tab:NILM}.
Beside the used appliances also the reached accuracy of the proposed load disaggregator is presented.
In this context, the \ac{ACC} is defined as:
\begin{equation}
ACC = \frac{TP+TN}{N},
\end{equation}
where $TP$ (true positives) is the number of times an appliance is correctly detected as ON, $TN$ (true negatives) is the number of times an appliance is correctly detected as OFF, while $N$ is the number of samples in the observation window.
The results in Table \ref{tab:NILM} show that the presented dataset can be applied as a reference dataset to load disaggregation problems.
The complexity of the problem can be adjusted according to the used appliances and corresponding appliance model (on/off device such as the water kettle or multi-state devices such as the dishwasher) and the number of aggregated power loads.
\begin{table}
\centering
\caption{Accuracy of load disaggregator for the aggregated power draws of houses \#0 and \#2}
 \rowcolors{1}{}{lightgray}
\begin{tabular}{l|c|c|cc}
\hline
\multicolumn{2}{c|}{\textbf{House ID 0}}&\multicolumn{2}{c}{\textbf{House ID 2}}\\
\hline
\textbf{Type}			& \textbf{ACC}	& \textbf{Type} 		& \textbf{ACC}\\
\hline
TV 						& 0.91			&	hair dryer			& 0.99\\
coffee machine  		& 0.99  	 	&	light 				& 0.97\\
dishwasher  			& 0.96		 	&	dishwasher			& 0.46\\
fridge  				& 0.93			&	fridge				& 0.90\\
vacuum cleaner			& 0.99			&	water kettle		& 0.99\\
water kettle  			& 0.99			&	washing machine		& 0.80\\
\hline
\end{tabular}
\label{tab:NILM}
\end{table}

\subsection{Occupancy detection} \label{sec:occupancy}
Occupancy detection is the problem of inferring presence of people in environments.
Many different approaches have been considered in the literature: motion detection, doors opening, use of acoustic sensors and/or cameras, use of GPS and localization systems through smart phones.
Nevertheless, only a few works propose to use energy consumption data to develop occupancy detection techniques \cite{Chen2013,Kleiminger2013}.
To this end, the assessment of occupancy detection requires suitable datasets, offering either an aggregated or a disaggregated power draw, along with a description of consumption scenarios.
In particular, the presence of user-driven devices is relevant to assume presence in the environment.
It is important to specify whether residents can postpone the operation of user-driven devices, as this will condition consumption-based approaches.
A simple approach based on the use of energy consumption data to detect occupancy was presented in \cite{Chen2013}.
Therein, the authors developed an algorithm called non-intrusive occupancy monitoring (NIOM).
In particular, NIOM guesses the presence of people at home by comparing the average, the variance and the maximum value of the current power consumption with threshold values.
The threshold values are computed during inactivity periods, namely when the activity of residents does not add power consumption to the baseline consumption, e.g.,
the consumption of devices such as fridge, HVAC systems etc.
Accordingly, the thresholds are computed every night when people are supposed to sleep.
As both power values and threshold values are computed over time slots of fixed duration, the duration of the slot is an important parameter affecting the estimation.
Although occupancy detection through energy consumption observation is a cheap and non-intrusive solution,
it is shown being rather inaccurate in determining the occupancy during periods of inactivity.

In order to give an example on the use of the GREEND dataset to study occupancy, we consider the house \#5.
We chose this scenario because we have measured the total power consumption, which allows for better estimating the baseline during inactivity periods.
We consider two months of measurements (February and March 2014) and we apply the NIOM algorithm as described in \cite{Chen2013}.
The time slot duration is chosen equal to 15 minutes. We consider two cases, the weekdays (WDs) and the weekend days (WEs).
This is because the analyzed power consumption greatly differs in these two circumstances as it can be seen in Fig. \ref{fig:NIOM_WD_energy} and Fig. \ref{fig:NIOM_WE_energy},
where the 99th percentile of the energy consumption during WDs and WEs are shown.
The baseline, within which the baseline threshold values are computed, was chosen at night between midnight and 6.30 AM.
Fig. \ref{fig:NIOM_WD} reports the occupancy obtained applying the NIOM algorithm over three consecutive days.
Fig. \ref{fig:NIOM_WD_prob} and Fig. \ref{fig:NIOM_WE_prob} respectively show the probability of occupancy obtained for WDs and WEs.
We first notice that, as expected, the results obtained between midnight and 6.30 AM do not indicate any activity.
We notice a good agreement between the derived probability of occupancy and the habits of residents.
Residents stated that during WDs, they usually wake up at 6.45 AM, leave around 8.00 AM to go to work and are back home around 6.00 PM.
During WEs residents tend to spend more time at home during the day and they go out at night coming back home around midnight.
\setlength\figureheight{2.6cm}
\setlength\figurewidth{7cm}

\begin{figure}[htp]
\centering
\subfloat[Occupancy over 3 days]{
	\input{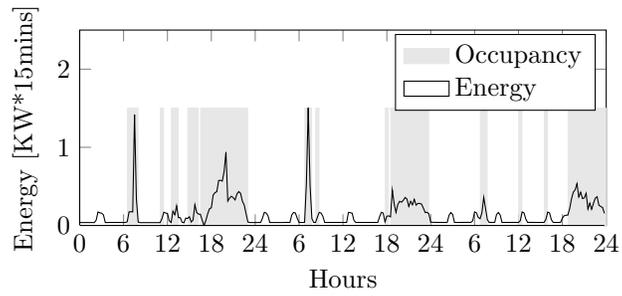}
	\label{fig:NIOM_WD}
}
\hfill
\subfloat[Weekday occupancy probability]{
	\definecolor{mycolor1}{rgb}{0.90588,0.90588,0.90588}%
\begin{tikzpicture}

\begin{axis}[%
width=\figurewidth,
height=\figureheight,
scale only axis,
xmin=0,
xmax=24,
xtick={0, 2, 4, 6, 8, 10, 12, 14, 16, 18, 20, 22, 24},
xlabel={Hours},
ymin=0,
ymax=1,
ylabel={Probability},
name=plot2
]
\addplot [color=black,solid,forget plot]
  table[row sep=crcr]{0	0\\
0.25	0\\
0.5	0\\
0.75	0\\
1	0\\
1.25	0\\
1.5	0\\
1.75	0\\
2	0\\
2.25	0\\
2.5	0\\
2.75	0\\
3	0\\
3.25	0\\
3.5	0\\
3.75	0\\
4	0\\
4.25	0\\
4.5	0\\
4.75	0\\
5	0\\
5.25	0\\
5.5	0\\
5.75	0\\
6	0\\
6.25	0.111111111111111\\
6.5	0\\
6.75	0.305555555555556\\
7	0.5\\
7.25	0.722222222222222\\
7.5	0.75\\
7.75	0.25\\
8	0.0277777777777778\\
8.25	0.0277777777777778\\
8.5	0.111111111111111\\
8.75	0.0555555555555556\\
9	0.138888888888889\\
9.25	0.0833333333333333\\
9.5	0.0555555555555556\\
9.75	0.111111111111111\\
10	0.0555555555555556\\
10.25	0.194444444444444\\
10.5	0.0833333333333333\\
10.75	0.111111111111111\\
11	0.0277777777777778\\
11.25	0.0277777777777778\\
11.5	0.0277777777777778\\
11.75	0.0555555555555556\\
12	0.0833333333333333\\
12.25	0.111111111111111\\
12.5	0.111111111111111\\
12.75	0.166666666666667\\
13	0.138888888888889\\
13.25	0.111111111111111\\
13.5	0.0555555555555556\\
13.75	0.0555555555555556\\
14	0.0277777777777778\\
14.25	0.0277777777777778\\
14.5	0.0555555555555556\\
14.75	0.0833333333333333\\
15	0.138888888888889\\
15.25	0.111111111111111\\
15.5	0.111111111111111\\
15.75	0.25\\
16	0.166666666666667\\
16.25	0.166666666666667\\
16.5	0.0555555555555556\\
16.75	0.166666666666667\\
17	0\\
17.25	0.0555555555555556\\
17.5	0.0833333333333333\\
17.75	0.111111111111111\\
18	0.416666666666667\\
18.25	0.75\\
18.5	0.75\\
18.75	0.805555555555556\\
19	0.861111111111111\\
19.25	0.805555555555556\\
19.5	0.75\\
19.75	0.722222222222222\\
20	0.722222222222222\\
20.25	0.777777777777778\\
20.5	0.861111111111111\\
20.75	0.916666666666667\\
21	0.861111111111111\\
21.25	0.888888888888889\\
21.5	0.888888888888889\\
21.75	0.861111111111111\\
22	0.888888888888889\\
22.25	0.833333333333333\\
22.5	0.833333333333333\\
22.75	0.805555555555556\\
23	0.694444444444444\\
23.25	0.444444444444444\\
23.5	0.277777777777778\\
23.75	0.305555555555556\\
};
\end{axis}
\end{tikzpicture}%
	\label{fig:NIOM_WD_prob}
}
\hfill
\subfloat[Weekday energy]{
	\input{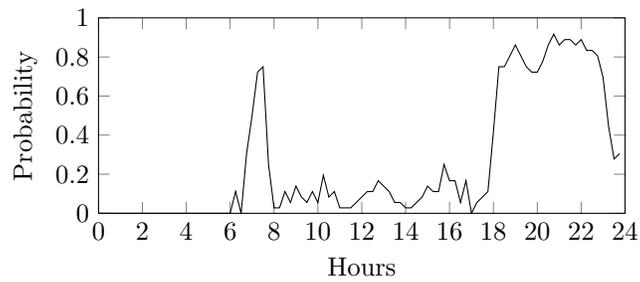}
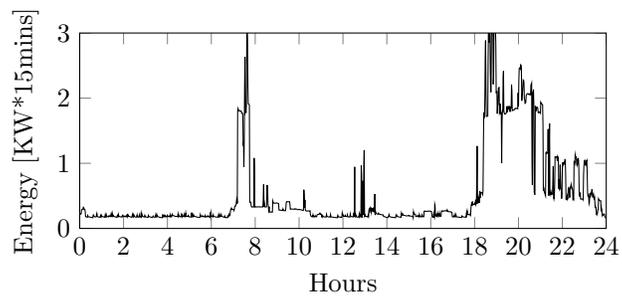
	\label{fig:NIOM_WD_energy}
}
\caption{Occupancy detection for weekdays in house \#5}
\label{fig:NIOM_WD}
\end{figure}

\begin{figure}[h!t]
\centering
\subfloat[Weekend occupancy probability]{
	\definecolor{mycolor1}{rgb}{0.90588,0.90588,0.90588}%
\begin{tikzpicture}

\begin{axis}[%
width=\figurewidth,
height=\figureheight,
scale only axis,
xmin=0,
xmax=24,
xtick={0, 2, 4, 6, 8, 10, 12, 14, 16, 18, 20, 22, 24},
xlabel={Hours},
ymin=0,
ymax=1,
ylabel={Probability},
name=plot2
]
\addplot [color=black,solid,forget plot]
  table[row sep=crcr]{0	0\\
0.25	0\\
0.5	0\\
0.75	0\\
1	0\\
1.25	0\\
1.5	0\\
1.75	0\\
2	0\\
2.25	0\\
2.5	0\\
2.75	0\\
3	0.0666666666666667\\
3.25	0.0666666666666667\\
3.5	0.0666666666666667\\
3.75	0.0666666666666667\\
4	0.0666666666666667\\
4.25	0.2\\
4.5	0.133333333333333\\
4.75	0.2\\
5	0.0666666666666667\\
5.25	0\\
5.5	0.133333333333333\\
5.75	0.133333333333333\\
6	0.133333333333333\\
6.25	0.133333333333333\\
6.5	0\\
6.75	0.2\\
7	0.2\\
7.25	0.0666666666666667\\
7.5	0.0666666666666667\\
7.75	0.0666666666666667\\
8	0.133333333333333\\
8.25	0.2\\
8.5	0.4\\
8.75	0.266666666666667\\
9	0.333333333333333\\
9.25	0.6\\
9.5	0.6\\
9.75	0.666666666666667\\
10	0.4\\
10.25	0.6\\
10.5	0.6\\
10.75	0.4\\
11	0.533333333333333\\
11.25	0.266666666666667\\
11.5	0.4\\
11.75	0.4\\
12	0.4\\
12.25	0.533333333333333\\
12.5	0.4\\
12.75	0.266666666666667\\
13	0.533333333333333\\
13.25	0.6\\
13.5	0.666666666666667\\
13.75	0.866666666666667\\
14	0.733333333333333\\
14.25	0.8\\
14.5	0.6\\
14.75	0.466666666666667\\
15	0.4\\
15.25	0.4\\
15.5	0.4\\
15.75	0.4\\
16	0.266666666666667\\
16.25	0.4\\
16.5	0.533333333333333\\
16.75	0.466666666666667\\
17	0.6\\
17.25	0.533333333333333\\
17.5	0.333333333333333\\
17.75	0.533333333333333\\
18	0.466666666666667\\
18.25	0.6\\
18.5	0.733333333333333\\
18.75	0.6\\
19	0.733333333333333\\
19.25	0.733333333333333\\
19.5	1\\
19.75	1\\
20	0.933333333333333\\
20.25	1\\
20.5	0.933333333333333\\
20.75	1\\
21	0.866666666666667\\
21.25	0.933333333333333\\
21.5	0.733333333333333\\
21.75	0.733333333333333\\
22	0.6\\
22.25	0.466666666666667\\
22.5	0.466666666666667\\
22.75	0.533333333333333\\
23	0.466666666666667\\
23.25	0.2\\
23.5	0.266666666666667\\
23.75	0.466666666666667\\
};
\end{axis}
\end{tikzpicture}%
	\label{fig:NIOM_WE_prob}
}
\hfill
\subfloat[Weekend energy]{
	\input{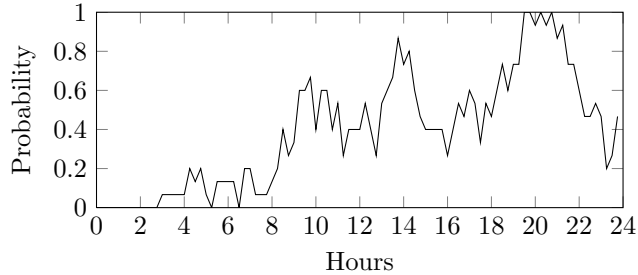}
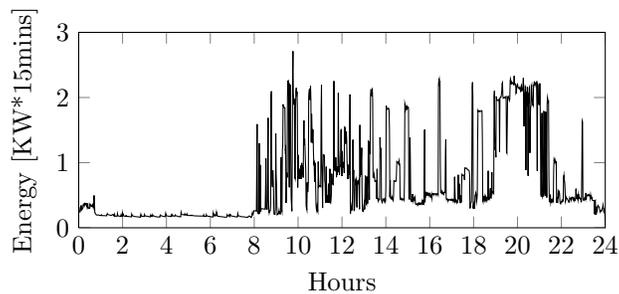
	\label{fig:NIOM_WE_energy}
}
\caption{Occupancy detection for weekends in house \#5}
\label{fig:NIOM_WE}
\end{figure}

\newpage
\subsection{Appliance usage modeling} \label{sec:usage}
Mining appliance usage patterns concerns the extraction of models describing how appliances are used by residents, given a sequence describing changes on their operational state.
%
Given log data describing status changes of electrical devices, the extraction of models describing the usage of devices can be achieved using different approaches, such as association rule mining \cite{Kang20121355}, artificial neural networks (ANNs) \cite{Aydinalp200287}, episode-generating Hidden Markov models (EGH) \cite{eps351238} and Bayesian networks \cite{Aguilera20111376}.
%
To show a possible application of the dataset, we propose to use a \ac{BN} to predict the usage of user-driven devices.
A \ac{BN} is a probabilistic graphical model encoding the joint probability distribution of a set of random variables.
A \ac{BN} $G = (V,E)$ is a directed acyclic graph (DAG) whose nodes $V = \{X_1, \ldots, X_n\}$ are random variables, while dependency between variable is represented by edges such as $E = X_i \rightarrow X_j$, and quantified by the conditional probability $P(X_j|X_i)$. 
In particular, each node is associated to a conditional probability distribution (CPD) quantifying the effect the parents have on the node, that is $P(X_1,\ldots, X_n) = \prod\limits_{i=1}^n P(X_i|pa(X_i))$ \cite{Koller:2009}.

\begin{figure}[h!tp]
	\centering
	\includegraphics[width=0.5\columnwidth]{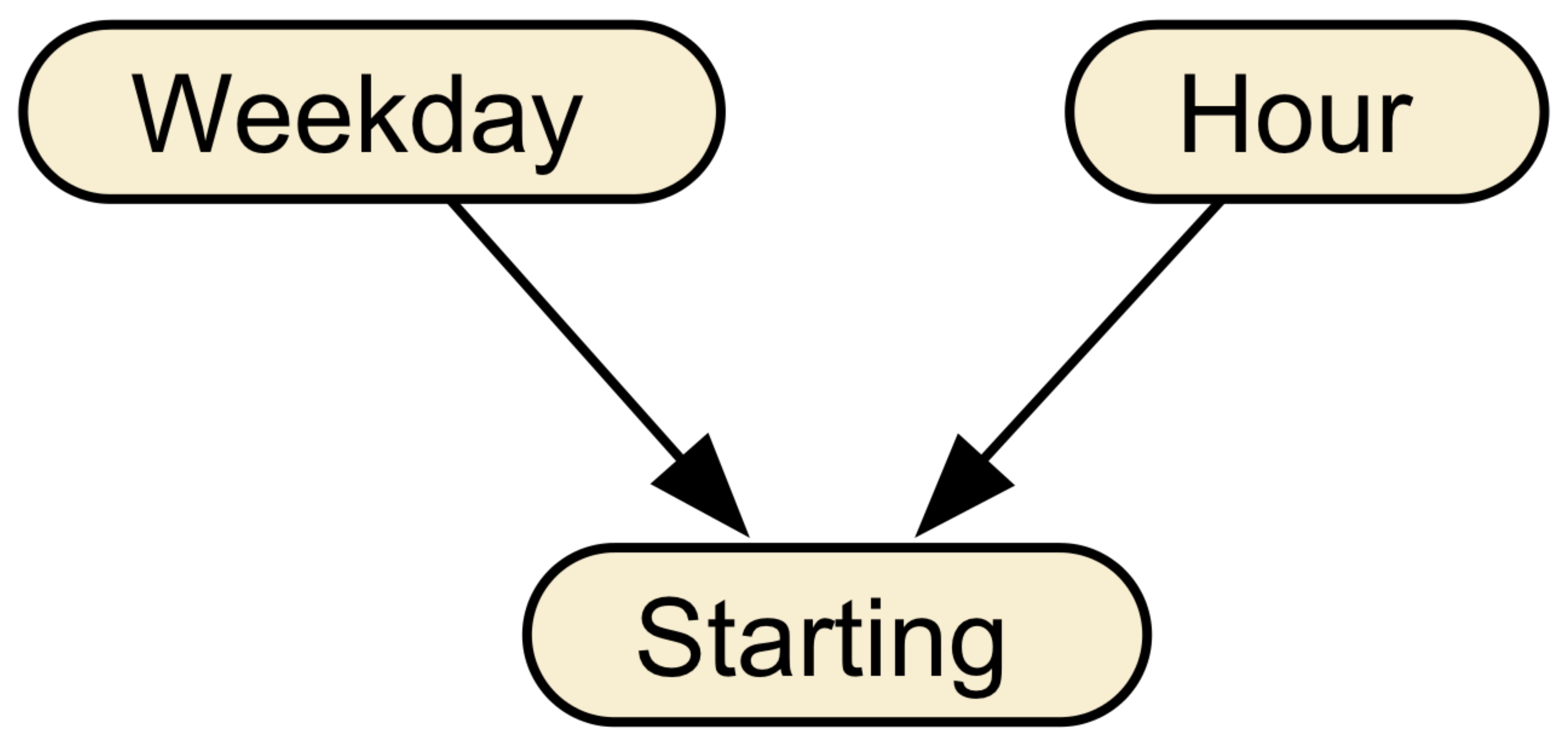}
	\caption{Bayesian network for the coffee machine of House \#0}
	\label{fig:network}
\end{figure}

The approach undertaken follows the work presented in \cite{bayesian_usage,drugan} from which we derived the network structure.
Because of the limited length of the dataset at time of writing we decided to omit seasonal information expressed through the month (see Fig. \ref{fig:network}).
As a first step, we processed the daily CSV files collected in the household ID\#0 and extracted starting events for each monitored device.
The \ac{BN} was implemented using the Netica\footnote{https://www.norsys.com} tool.
In particular, we used \ac{EM} to perform parameter learning for the given network and we used the \ac{JT} algorithm to perform exact inference.
Netica provides APIs for various programming languages, making the learned model applicable in working solutions.
The plot in Fig.\ref{fig:startingprob} reports the posterior probability of starting the coffee machine of household ID\#0, given the observations day of the week (i.e. weekday, Saturday and Sunday) and hour of the day.
As noticeable, the patterns of coffee making tend to be regular over the day.
The modeled couple tends to wake up earlier during weekdays, which was confirmed by an interview with the householders.
\setlength\figureheight{5.5cm}
\setlength\figurewidth{7cm}
\begin{figure}[h!tp]
	\centering
%
%
%
\definecolor{mycolor1}{rgb}{0.24706,0.24706,0.24706}%
\begin{tikzpicture}

\begin{axis}[%
width=\figurewidth,
height=0.85\figureheight,
scale only axis,
xmin=0,
xmax=24,
xtick={ 0,  4,  8, 12, 16, 20, 24},
xlabel={Hour},
ymin=0,
ymax=1,
ytick={0,0.2,0.4,0.6,0.8,1},
yticklabels={{0},{0.2},{0.4},{0.6},{0.8},{1}},
ylabel={Probability to Start},
legend style={draw=black,fill=white,legend cell align=left,legend pos=north west}
]
\addplot [color=black,solid,line width=1.0pt,mark=triangle*, mark options={fill=white}]
  table[row sep=crcr]{1	1.99999e-06\\
2	1.99999e-06\\
3	1.99999e-06\\
4	1.99999e-06\\
5	1.99999e-06\\
6	1.99999e-06\\
7	1.99999e-06\\
8	1.99999e-06\\
9	0.6\\
10	0.799999\\
11	0.4\\
12	0.4\\
13	1.99999e-06\\
14	1.99999e-06\\
15	0.4\\
16	0.200001\\
17	0.200001\\
18	0.4\\
19	0.200001\\
20	0.200001\\
21	1.99999e-06\\
22	1.99999e-06\\
23	1.99999e-06\\
24	1.99999e-06\\
};
\addlegendentry{SAT};

\addplot [color=black,dashed,line width=1.0pt,mark=square*, mark options={solid,fill=white}]
  table[row sep=crcr]{1	1.99999e-06\\
2	1.99999e-06\\
3	2.49999e-06\\
4	1.99999e-06\\
5	1.99999e-06\\
6	1.99999e-06\\
7	1.99999e-06\\
8	0.200001\\
9	0.6\\
10	0.799999\\
11	0.799999\\
12	0.5\\
13	0.5\\
14	0.5\\
15	0.5\\
16	0.333334\\
17	0.5\\
18	0.666666\\
19	0.5\\
20	0.166668\\
21	1.66666e-06\\
22	1.66666e-06\\
23	1.66666e-06\\
24	1.66666e-06\\
};
\addlegendentry{SUN};

\addplot [color=black,dotted,line width=1.0pt,mark=+,mark options={solid}]
  table[row sep=crcr]{1	3.84615e-07\\
2	3.84615e-07\\
3	3.84615e-07\\
4	3.84615e-07\\
5	3.84615e-07\\
6	3.84615e-07\\
7	0.5\\
8	0.769231\\
9	0.923077\\
10	0.961538\\
11	0.807692\\
12	0.576923\\
13	0.153846\\
14	0.423077\\
15	0.576923\\
16	0.615385\\
17	0.538462\\
18	0.269231\\
19	0.269231\\
20	0.0769234\\
21	0.0384619\\
22	0.0400004\\
23	0.0400004\\
24	4e-07\\
};
\addlegendentry{WD};

\end{axis}
\end{tikzpicture}%
	\caption{Usage forecasting for the coffee machine of House \#0}
	\label{fig:startingprob}
\end{figure}
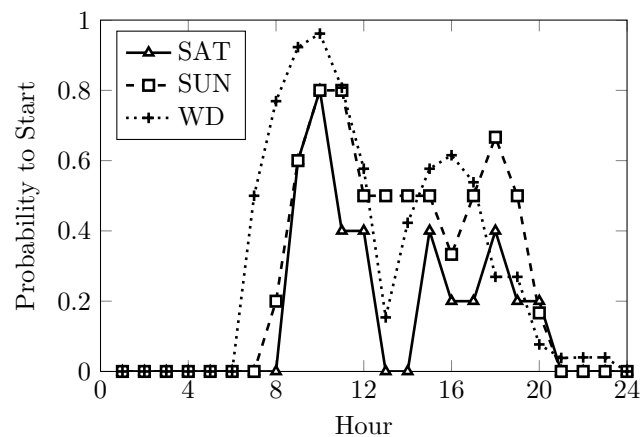
\section{Conclusions and future work} \label{sec:conclusion}
We have presented the GREEND consumption dataset, obtained through a measurement campaign in households in Austria and Italy.
The dataset is available for open use and consists of consumption data of selected devices.
Along with the dataset we have also released our code base for the campaign as a Sourceforge project.
This includes the scripts to collect consumption data and a ready-to-use SD image for the Raspberry Pi, as well as
processing scripts that can be used to extract edges and consumption events, based on configurable device-specific thresholds.
We expect our contribution to be of value for researchers and engineers dealing with domestic energy management systems.
To this end, we showed the use of GREEND to develop and assess techniques for load disaggregation, occupancy detection and appliance usage mining.
We expect to extend the monitoring campaign to collect aggregated consumption data, as this allows for modeling the whole demand, which is a valuable information when emulating microgrids and assessing control strategies at a wider scope.
Similarly, we plan to include data concerning production from renewable energy, as it can be of value to forecast energy generation.
The presence of temporal information, along with the locality and therefore the weather, will allow for the extraction of seasonal consumption patterns.
The long term measurement campaign carried out within MONERGY will provide the time span to allow this type of studies.
We expect to ultimately provide ready-to-use models of the regions, that can be imported in current simulation tools for smart grid applications.
\bibliographystyle{IEEEtran}
\bibliography{NIALM}
\end{document}